\documentclass[pss]{wiley2sp} % provides pss two-column style
\usepackage{amsmath}
\usepackage{bm}
\usepackage{amsfonts}
\usepackage[T1]{fontenc}

\newcommand{\E}{\mathbf{E}}%
\newcommand{\B}{\mathbf{B}}%
\newcommand{\A}{\mathbf{A}}%
\begin{document}

% Title of the article
\title{Topological Magnetoelectric Effect:\\
Nonlinear Time-Reversal-Symmetric Response, Witten Effect, and Half-Integer Quantum Hall Effect}

% Authors
\author{%
  Heinrich-Gregor Zirnstein\textsuperscript{\Ast,\textsf{\bfseries 1}},
  Bernd Rosenow\textsuperscript{\textsf{\bfseries 1}}}

%E-mail-address of corresponding author
\mail{e-mail
  \textsf{zirnstein@itp.uni-leipzig.de}}

% author's affiliations/addresses
\institute{%
  \textsuperscript{1}\,Institut f\"ur Theoretische Physik, Universit\"at Leipzig, D-04103 Leipzig, Germany}

% Please select about four verbal keywords for your manuscript.
\keywords{topological insulator, magnetoelectric effect, axion electrodynamics, quantum Hall effect}

\abstract{\bf%
Topological insulators (TIs) in three space dimensions
can be characterized by a quantized magnetoelectric coefficient. However, this coupling does not have experimentally observable consequences in the presence of time-reversal symmetry, because the contributions of both bulk and surface states cancel each other. Instead, the characteristic response of a TI is a \emph{nonlinear} magnetoelectric effect. We discuss both field theoretic aspects of the nonlinear magnetoelectric response and numerical calculations for experimentally relevant geometries. Distinct from this effect,  the magnetoelectric coupling would bind a charge $\pm e/2$ in response to a monopole, which is referred to as Witten effect. If time reversal is broken explicitly, for instance by a Zeeman field acting on the surface layer, the electromagnetic response of a TI can be described by a half-integer quantum Hall effect of surface electrons.
}

\maketitle   % please do not remove

%%%%%%%%%%%%%%%%%%%%%%%%%%%%%%%%%%%%%%%%%%%%%%%%%%%%%%%%%%%%
\section{Introduction}

Topological insulators (TIs) are electronic materials that are insulating (gapped) in the bulk, but feature conducting (gapless) surface states~\cite{Hasan:2010a,Qi:2011,Chiu:2016a}.
% assuming that electron interactions can be ignored. 
These surface states are topologically protected, which means that they are robust against  perturbations of the system parameters, as long as bulk gap is not closed and the material stays insulating. An additional symmetry may be required, for instance time-reversal (TR) symmetry.

Time-reversal-symmetric TIs in three space dimensions feature gapless surface states whose energy dispersion is described by two-dimensional Dirac cones~\cite{Hasan:2010a,Qi:2011}. The low-energy properties of the bulk have been described by the so-called axion action, which has been associated with a topological magnetoelectric effect, where an electric field induces a magnetic field and vice versa~\cite{Qi:2008,Essin:2009a,Essin:2010}. However, a naive interpretation of the axion action as a material response would imply a paradoxical breaking of time-reversal symmetry.
In this Feature Article, we summarize the status of the topological magnetoelectric effects in TIs. In particular, we clarify that for the experimental response, both bulk and surface contributions have to be taken into account, and their contributions cancel each other ~\cite{Mulligan:2013,Zirnstein:2013a}, and as a consequence   the linear magnetoelectric effect vanishes.
Any experimentally observable magnetoelectric response has to be \emph{nonlinear} for this reason.
Current experiments~\cite{Chang:2013,Xu:2014,Wu:2016a,Okada:2016,Dziom:2017} focus on the response for a system where time-reversal symmetry has been broken explicitly, e.g.~by a strong magnetic field or a Zeeman field. For this setup, the signature response is a quantum Hall effect, which is \emph{half}-integer for the surface states of the TI, instead of integer as one would expect for an ordinary two-dimensional electron gas.
In contrast, for a time-reversal-symmetric setup, the signature response is a nonlinear magnetoelectric effect ~\cite{Zirnstein:2017}, which in the presence of a small electric field leads to the appearance of half-integer charges bound to a flux quantum, see Fig.~\ref{fig:effect}. 
The physics discussed in this paper is not related to the wormhole effect~\cite{Rosenberg:2010}, which only occurs for lattice models where the magnetic flux is highly focused and fits through a single lattice plaquette. Instead, we consider electromagnetic fields that extend over several lattice spacings, so that a continuum description is applicable.

%%%%%%%%%%%%%%%%%%%%%%%%%%%%%%%%%%%%%%%%%%%%%%%%%%%%%%%%%%%%
\section{Apparent Linear Magnetoelectric Effect}

A well-known example for a topological state of matter in two space dimensions 
is the integer quantum Hall effect: Even though the bulk of a quantum Hall state is
insulating, such that the longitudinal conductance vanishes, there is a nonzero Hall current $\mathbf{j}_{\perp}$ which flows perpendicular to an
applied electric field $\E_{\parallel}$. The Hall conductance, which
relates the two via $\mathbf{j}_{\perp}=\sigma_{xy}\E_{\parallel}$ is quantized
to an integer multiple of the conductance quantum $e^{2}/h$. Importantly,
this integer ($\mathbb{Z}$) is equal to the topological invariant
characterizing the topological insulator, thus enabling us to determine
the invariant with an experimental measurement~\cite{Hasan:2010a,Haldane:1988}.

In three space dimensions, one might also hope that the electromagnetic
response of the bulk material can be related to the topological invariant.
In this case, the invariant is not integer-valued, but takes one of
two values (``even'' or ``odd'', $\mathbb{Z}_{2}$). Indeed, it
has been proposed~\cite{Qi:2008,Essin:2010} that time-reversal-symmetric
TIs are characterized by a magnetoelectric effect of topological origin,
where applying an external electric field will induce a magnetic field in response.
It can be described by adding the so-called \emph{axion action}
\begin{equation}
  S_{\theta} = \frac{\theta}{2\pi}\frac{e^{2}}{2\pi\hbar c} \int d^{3}xdt\,\E\cdot\B
\label{eq:axion}
\end{equation}
to the standard Maxwell action. Here, $\theta$ is  the value of the
topological invariant: $\theta=0$ for a trivial insulator (``even''),
and $\theta=\pm\pi$ for a topological insulator (``odd''). We use the conventions $\hbar=1$ and $c=1$ from now on.~\footnote{We also use Gaussian units throughout.}

Unfortunately, such a linear magnetoelectric effect seems to be at odds with time-reversal symmetry: If an electric field produces a magnetization by inducing currents in the material, in which direction do these currents flow? Reversing the direction of time would change their direction, but symmetry requires them to be unchanged, so they have to vanish.
More formally, we can look at the action Eq.~(\ref{eq:axion}): A time-reversal transformation maps $\E \to \E$, $\B \to -\B$, and hence $S_{\theta}\to-S_{\theta}$.
We conclude that the action is unchanged only if it vanishes, which means $\theta = 0$.
However, there is a loophole: In quantum mechanics, it is not the classical action that needs to be invariant, but the exponential $e^{iS_{\theta}}$, because only the latter appears in a Feynman path integral. The latter is still invariant if $S_\theta$ is quantized to be an integer multiple of $\pi$. Indeed, it turns out that for periodic boundary conditions, the integral over $\E\cdot\B$ itself is quantized to an integer multiple of $4\pi^{2}/e^{2}$~\cite{Milnor:1974,Vazifeh:2010}. This quantization is a quantum mechanical effect, and follows from the requirement that it must be possible to consistently couple an electronic wave function to the electromagnetic gauge field. In any case, this quantization implies $S_{\theta}=\theta\mod2\pi$, and we conclude that $\theta=\pm\pi$ would also be compatible with time-reversal symmetry. This is indeed the value realized in a topological insulator.

What does this mean for the material response?
The previous argument already implies that the axion action $S_{\theta}$ is rather unusual. In particular, we have just argued that for periodic boundary conditions, the action is quantized, i.e.~constant. But in classical physics, the response is given by the Euler-Lagrange equations, which are obtained by variation of the fields. But since the action is constant, its variation is zero, and it contributes nothing to the response. Thus, for periodic boundary conditions, the axion action Eq.~(\ref{eq:axion}) appears to have no effect on the electromagnetic response.
More generally, the existence of a magnetoelectric effect implies that the magnetic field induces an electric polarization, and the electric field a magnetization~\cite{Landau:1984}, so that the constituent relations of the material are modified to read
\begin{subequations}
\label{eq:constituent}
\begin{align}
  \mathbf{D} &= \varepsilon \E - 4\pi\alpha\B \ \ 
,\\
  \mathbf{H} &= \mu^{-1}\B + 4\pi\alpha \E \ \ 
  \label{eq:constituentH}
,\end{align}
\end{subequations}
where $\alpha = \frac{\theta}{2\pi}\frac{e^{2}}{2\pi\hbar c}$ is the magnetoelectric coefficient~\cite{Qi:2008,Qi:2009}.
But if we use these relations in the macroscopic Maxwell-Amp\`{e}re law
$\nabla \times \mathbf{H} = 4\pi \mathbf{j}_{\text{ext}} + \partial_t \mathbf{D}$, where $\mathbf{j}_{\text{ext}}$ denotes the external current, and use the law of induction, $\nabla\times\E=-\partial_t \B$, we find that the terms involving the magnetoelectric coefficient $\alpha$ cancel, and we are left with the standard Maxwell equation
\begin{equation}
  \nabla \times (\mu^{-1} \B)
  =
  4\pi\mathbf{j}_{\text{ext}} + \frac{\partial}{\partial t} (\varepsilon \E) \ \ 
.\end{equation}
In other words, it is indeed true that a constant magnetoelectric coefficient $\alpha$ in a material with periodic boundary conditions does not contribute anything to the response. This state of affairs is time-reversal-symmetric, \emph{regardless} of the value of $\theta$.

The discussed lack of a response from  the axion action is due to periodic boundary conditions. In contrast, experiments are done with finite materials, which always have boundary surfaces. In this case, the magnetoelectric coefficient $\alpha$ does lead to a classical response, because it is nonzero within the material, but zero outside.\footnote{More generally, spatial variation of $\alpha$ leads to a nonzero response; this can be seen by using Eq.~\eqref{eq:constituent} in the macroscopic Maxwell-Amp\`{e}re law, while allowing $\alpha\equiv\alpha(\mathbf{x})$ to vary in space.~\cite{Armitage:2019}}
It turns out that the response due to a spatial change in $\alpha$ is nonzero only on the surface of the TI, and is equivalent to a half-integer Hall effect. This result is due to the fact that  the axion action Eq.~(\ref{eq:axion}) can  be transformed into a surface integral.
To see this, we need to express the electromagnetic field in terms of the electric potential $A_0$ and the magnetic vector potential $\mathbf{A}$, such that $\mathbf{B}=\nabla\times\mathbf{A}$ and $\mathbf{E}=-\nabla A_0 - \partial_t \mathbf{A}$. In fact, the calculation becomes more transparent if we use the electromagnetic field tensor $F_{\mu\nu} = \partial_\mu A_\nu - \partial_\nu A_\mu$; see Appendix \ref{sec:electrodynamics} for the standard relations. 
We find~\cite{Qi:2009,Armitage:2019}
\begin{align}
  \nonumber
S_{\theta} & = \frac{\alpha}{8}
  \int_{V}d^{3}x\int dt\,\varepsilon^{\alpha\beta\gamma\delta}F_{\alpha\beta}F_{\gamma\delta}\\
\nonumber
  & = \frac{\alpha}{8}
  \int_{V}d^{3}x\int dt\,\varepsilon^{\alpha\beta\gamma\delta}(\partial_{\alpha}A_{\beta}-\partial_{\beta}A_{\alpha})F_{\gamma\delta}\\
  \nonumber
 & = \frac{\alpha}{4} \int_{V}d^{3}x\int dt\,\left[\partial_{\alpha}(\varepsilon^{\alpha\beta\gamma\delta}A_{\beta}F_{\gamma\delta})-\varepsilon^{\alpha\beta\gamma\delta}A_{\beta}\partial_{\alpha}F_{\gamma\delta}\right]\\
 &= \frac{\alpha}{4} \int_{\partial V}d\mathbf{n}_{a}\int dt\,\varepsilon^{a\beta\gamma\delta}A_{\beta}F_{\gamma\delta}.
\end{align}
Here, $V$ denotes the spatial region occupied by the material, $\partial V$
its boundary, and $\mathbf{n}_{a}$ its normal vector. We have discarded the integral over the temporal boundary because it does not influence the response. Also, $\varepsilon^{\alpha\beta\gamma\delta}$ is the fully antisymmetric tensor and we have used that the divergence of the field strength vanishes, $\varepsilon^{\alpha\mu\nu}\partial_{\alpha}F_{\mu\nu}=0$.
The remaining surface integral describes a Hall response. For instance, if we specialize to the half-space below the plane $x_{3}=0$, then the action becomes
\begin{equation}
S_{\theta}= -\frac{\alpha}{4} \int_{x_{3}=0}d^{2}x\int  dt\,\varepsilon^{\alpha\beta\gamma}A_{\alpha}F_{\beta\gamma}
.\end{equation}
The current is given by the functional derivative
\begin{subequations}
\begin{align}
j^{1} & = \frac{\delta S}{\delta A_{1}} = \alpha F_{20}= \alpha\E_{2} \ \ 
,\\
j^{2} & = \frac{\delta S}{\delta A_{2}} = \alpha F_{01}= -\alpha\E_{1} \ \ 
,\end{align}
\end{subequations}
such that the current is proportional to the applied electric field, and
perpendicular to its direction. In conclusion, the axion action corresponds to a Hall current with Hall conductivity $\sigma_{xy}=\alpha$ on the surface.

%%%%%%%%%%%%%%%%%%%%%%%%%%%%%%%%%%%%%%%%%%%%%%%%%%%%%%%%%%%%
\section{Anomaly matching}

We have seen above that for a finite system with boundary, the axion action corresponds to a quantum Hall response on the surface. For the value $\theta = \pm \pi$, the Hall conductivity is half the conductance quantum, $\sigma_{xy} = \pm (1/2)(e^2/h)$.
However, this response clearly breaks time-reversal symmetry.
In contrast to the case with periodic boundary conditions, the axion action is no longer quantized, and the previous argument why this particular value of $\theta$ should be compatible with time-reversal symmetry fails.

The response of a time-reversal-symmetric system always has to be time-reversal-symmetric (unless the symmetry is spontaneously broken, which does not happen for noninteracting electrons), and this applies to topological insulators as well. 
For this reason, the idea that  the axion action describes the response of a finite time-reversal-symmetric topological insulator is a misconception and is incorrect.
Instead, in the presence of boundaries, the topological insulator features topologically protected gapless states localized at the boundary, whose response is \emph{not included} in the axion action, but \emph{must be taken into account} to obtain the total response.
This statement is correct because the axion action has been calculated for periodic boundary conditions~\cite{Essin:2010}, and thus only describes the response of the bulk states.
The fact that after partial integration it describes a response localized on the surface may be unusual, but does not contradict the fact that it describes the 
response of bulk states.
Moreover, since the total electromagnetic response of the TI must be time-reversal-symmetric, the only logical possibility is that the response due to boundary states also breaks time-reversal symmetry, such that  that the total response of bulk and boundary together is again time-reversal-symmetric. Indeed, this has been verified explicitly in Ref.~\cite{Mulligan:2013}.

It may seem strange that the electromagnetic response contributed by the surface states breaks time-reversal symmetry, even though the energy dispersion of surface states, a  Dirac cone, is time-reversal-symmetric. 
For a system with two surfaces, it has been argued that such a system  is invariant under large gauge transformations, because the contributions of the two surfaces cancel~\cite{Mulligan:2013}.
However, for a semi-infinite system with only one surface the answer to this puzzle is deep: The energy dispersion describes the motion of single particles, but calculating the electromagnetic response of the surface states is a many-particle problem, described by a quantum field theory. Since the energies in a Dirac cone are unbounded both above and below the Fermi level, one encounters divergent integrals, and one has to use regularization to make them finite. This procedure may break symmetries that are present in the classical action (dispersion relation), an effect known as a \emph{quantum anomaly}.
The fact that the whole system is time-reversal-symmetric corresponds to the requirement that the quantum anomaly of the surface states has to cancel with another quantum anomaly generated by other degrees of freedom. Indeed, the bulk states also have a quantum anomaly, described by the axion action, which must match and cancel the anomaly of the Dirac fermions on the surface~\cite{Zirnstein:2013a,Witten:2016a,Witten:2019}.
Anomaly matching has become a powerful principle for studying symmetry-protected topological phases of matter~\cite{Witten:2016a,Gaiotto:2017a}, as it is insensitive to interactions.
In fact, if we run the logic in reverse, the bulk anomaly predicts the presence of boundary states, because they are the only way for the total system to be symmetric.
In a related way, anomalies can be considered as an obstruction for regularizing a continuum theory on the lattice. In particular, this implies that the surface states of a topological insulator cannot be realized in a purely two-dimensional tight-binding model where the symmetry is on-site, they only exist as the low energy states of a three-dimensional bulk~\cite{Huang:2019}. An analogous example in odd space dimensions is provided by the Nielsen-Nimoniya theorem~\cite{Nielsen:1981,Friedan:1982}, also known as fermion-doubling theorem, which states that in a three-dimensional tight-binding model with on-site chiral symmetry, Weyl cones may only occur in pairs of opposite chirality. Thus, isolated Weyl cones may only be found as low energy states of a four-dimensional bulk.

In view of the above, one could object that in a lattice model, all integrals are finite, and the issue of regularization does not occur. The reply is that in a three-dimensional lattice, it is no longer possible to uniquely separate the surface states from the bulk states: At sufficiently high  momenta, the energies of the surface Dirac cone will become larger than the bulk energy gap, and the surface states will hybridize with the bulk. Thus, describing the surface with a Dirac cone is an approximation that is valid only below a certain energy cutoff, and difficulties are to be expected when calculating the contributions of higher energies to the response. If anything, the surprise here is that states at high energies, the bulk states, do contribute to the electromagnetic response at low energies. This seems to contradict the common wisdom that no currents can flow in a material if there are no electronic states at the corresponding energy, but the point is that this is only true for the longitudinal current; the Hall current, for instance, can still be nonzero.
We also note that when performing a numerical calculation of the response
for a finite lattice, say by diagonalizing the Hamiltonian, then all
states are automatically taken into account, regardless of whether they are surface,
bulk, or hybridized states.

%%%%%%%%%%%%%%%%%%%%%%%%%%%%%%%%%%%%%%%%%%%%%%%%%%%%%%%%%%%%
\section{Witten Effect}

So far we have discussed that in a system with periodic boundary conditions, the axion action generates no classical response, while in an open system, its effect is  canceled by the response of the topologically protected boundary states. Aside from its role in anomaly matching, does the axion action have any physical consequences in the form of a magnetoelectric bulk effect at all? The answer is affirmative, in form of the so-called Witten effect~\cite{Witten:1979,Rosenberg:2010a}, which however requires a quantum mechanical discussion together with  the existence of magnetic monopoles.

Historically, the axion action came to prominence in quantum field theory, where it describes a violation of $CP$-symmetry (charge-parity), or equivalently, of time-reversal symmetry $T$~\cite{tHooft:1976,Wilczek:1978}. In this context, Witten has shown that if we were to assume the existence of magnetic monopoles, then the axion action has the unusual consequence that these monopoles must carry an additional charge $e\theta /2\pi$~\cite{Witten:1979}. Using this argument for a time-reversal-symmetric TI with $\theta=\pm\pi$ would imply that bringing a magnetic monopole into the bulk of the material would lead to an accumulation of charge $\pm e/2$ at the monopole. This is indeed a bulk signature of the axion action, and has been termed the Witten effect~\cite{Rosenberg:2010a}.

To give a heuristic argument in favor of a charge response to a magnetic monopole, we appeal to the classical action. For static fields, we can write $\mathbf{E}=-\nabla A_0$. Using partial integration, we have
\begin{align}
S_{\theta}
\nonumber &=
  \frac{\theta e^2}{4\pi^{2}}\int d^{3}x dt\,\mathbf{E}\cdot\B
=
  \frac{\theta e^2}{4\pi^{2}}\int d^{3}x dt\,(-\nabla A_{0})\cdot\B
\\ &=
  \frac{\theta e^2}{4\pi^{2}}\int d^{3}x dt\,A_{0}(\nabla\cdot\B)
,\end{align}
where we have discarded the boundary term.
% see Appendix \ref{sec:electrodynamics}
We can calculate the charge density by performing the functional derivative $j^0 = -\delta S_\theta / \delta A_0$, and find that it is proportional to the divergence of the magnetic field, $j^0 \propto \nabla\cdot\B$, which is the monopole density.
Appealing to Dirac's quantum mechanical argument~\cite{Dirac:1931} that the magnetic charge $g$ and the elementary charge $e$ are related by the quantization condition $eg=2\pi n$ for integer $n$, we find that a monopole inside the material induces the charge
\begin{equation}
  q_{\theta} = \frac{e \theta }{2\pi}
\end{equation}
in the material, where we have assumed $n=-1$. For a topological insulator, $\theta = \pm \pi$, and this charge is half-integer $\pm e/2$.
That said, this derivation can only be  heuristic in nature, because magnetic monopoles cannot be modeled with a globally smooth vector potential $A_\mu$, as the very condition $\nabla\cdot \B=0$ is necessary to express the magnetic field as a curl, $\B = \nabla\times\A$.
\footnote{The use of the gauge field $A_{\mu}$ leads to another subtlety: The $\theta$-term was derived for noninteracting electrons on a lattice with periodic boundary conditions~\cite{Essin:2010}. In this setting, the electromagnetic field is introduced via minimal coupling, where each link in the lattice acquires a phase factor $\exp(i\int\mathbf{A}\cdot d\mathbf{l})$, which is determined by the line integral of the gauge field. Then, to obtain the response, one performs a gradient expansion in the gauge field. However, this approach is unsuitable for describing magnetic monopole configurations: Exponentiation means that the system is insensitive to adding one flux quantum $\Phi_{0}$ to a lattice plaquette, which, however, corresponds to adding more magnetic monopoles. Moreover, the gradient expansion needs to be adapted to the singularity at the monopole. Thus, strictly speaking, it is unclear whether this derivation of the $\theta$-term gives correct results in the presence of monopoles.}

While we refer to Witten~\cite{Witten:1979} for an accurate derivation of the effect in quantum field theory, there is another argument for the monopole charge $\pm e/2$ that is useful in the context of condensed matter.
For this, let us consider a finite topological insulator with boundaries. For the sake of the argument, we also assume that we can switch on an additional Zeeman field on the surface, which leads to a quantum Hall effect (to be discussed in more detail in the next section).
We imagine a thin magnetic flux tube carrying one flux quantum $\Phi_0$, which begins at infinity, but is terminated by a single magnetic monopole. First, the monopole is outside the topological insulator, and no magnetic flux passes through the material. Then, we switch on the Zeeman field and move the monopole to the inside of the material, so that one flux quantum pierces the surface. This process can be performed adiabatically, since both surface and bulk are gapped. Due the surface quantum Hall effect generated by the Zeeman field, the flux tube attracts a charge $\pm e/2$ when the monopole passes through; here, the sign of the charge depends on the sign of the Zeeman field. But since the total charge in the system is conserved, the monopole must have attracted an opposite charge $\mp e/2$. Switching off the Zeeman field again will return the system to a time-reversal-symmetric state, except that now a monopole is located inside it, which carries half-integer charge.

In this sense, the axion action binds a charge to a magnetic monopole in the bulk. However, this is not a classical response, in that the sign of this charge depends on the details of the adiabatic process by which the monopole is moved inside. It is more helpful to think of the monopole as inducing localized electronic states that can carry charge $\pm e/2$, but whether they are occupied or not in the many-body quantum state describing the electrons in the material will depend on additional circumstances. We will return to this point of view  shortly in the discussion of a time-reversal symmetric magnetoelectric effect.

However, in the end, as magnetic monopole have never been found as elementary particles, the Witten effect is not relevant for experiments.

%%%%%%%%%%%%%%%%%%%%%%%%%%%%%%%%%%%%%%%%%%%%%%%%%%%%%%%%%%%%
\section{Half-integer quantum Hall effect}

As discussed above, in the presence of time-reversal symmetry, topological insulators do not display a linear magnetoelectric effect. To find an experimental signature, one has to go beyond linear response, and study nonlinear effects in the electromagnetic response. For instance, one could continue the Taylor expansion of the action to orders beyond quadratic, obtaining terms like $\E^2\B^2$~\cite{Malshukov:2013}. However, a more popular choice is to consider the situation where some fields are large, so that the response saturates and becomes quantized.
In particular, in the presence of a strong external magnetic field, we expect that the  surface states exhibit a quantum Hall effect. %, similar to that of a regular  two-dimensional electron gas.
While the quantized Hall effect can be understood as a linear response, we emphasize that it is linear only if one neglects the magnetic field dependence of the Hall conductivity, i.e.~in a system where time-reversal symmetry is broken strongly and explicitly.

The key signature for Dirac fermion surface states is that their Hall conductivity is quantized as a \emph{half} integer multiple of the conductance quantum $e^2/h$, e.g.~$\sigma_{xy}=\pm (1/2)(e^2/h)$ in the presence of a strong orbital magnetic field. In contrast, the Hall conductivity of an electron gas obtained by confining the charge carriers in the third dimension, e.g.~in GaAs/AlGaAs heterostructures, is an integer multiple of the conductance quantum. This is intimately connected to the fact that Dirac fermion surface states can only be realized on the boundary of a three-dimensional atomic lattice, not in a two-dimensional lattice.

Several experimental consequences of the half-integer Hall effect have been observed or proposed. First, the Hall resistance of a sample in  slab geometry has been measured in transport experiments~\cite{Xu:2014}. However, since in this case both  top and bottom surfaces contribute, the total Hall resistance is found to be integer, and it is nontrivial to argue that each surface contributes a half-integer effect~\cite{Lee:2009,Xu:2014}.
It is also possible to detect the Hall conductivity in a contactless measurement  by employing electromagnetic radiation: The polarization planes of reflected (Kerr effect) and transmitted waves (Faraday effect) are rotated, because angular momentum is transferred to the electron system on the surfaces~\cite{Tse:2010,Maciejko:2010}. The above effects have recently been observed with THz radiation~\cite{Wu:2016a,Dziom:2017}.
Again, the contributions of two surfaces have to be taken into account when the radiation passes through a topological insulator layer~\cite{Wu:2016a}.
Another proposal, which would be able to detect the half-integral conductivity for a single surface, is the ``magnetic mirror monopole''~\cite{Qi:2009}, where bringing an electric point charge close to the TI will create currents on the surface whose magnetic field looks as if it were generated by a magnetic point charge inside the material. Of course, there is no actual monopole that corresponds to this mirror image, the divergence of the magnetic field still vanishes everywhere~\cite{Nogueira:2018}. This proposal has not yet been realized experimentally.

So far, we have discussed the quantum Hall effect obtained by subjecting the sample to a strong orbital magnetic field. However, an alternative method to break time-reversal symmetry is to introduce magnetic order on the surface. Theoretically, this can be modeled by a magnetic Zeeman field acting on the surface electrons, which corresponds to a mass term in the surface Dirac equation. In fact, careful reading reveals that one of the first derivations of the magnetoelectric response, by Qi et.al.~\cite{Qi:2008}, actually considers such a setup. The resulting Hall effect still has half-integer conductance, and it is often called \emph{quantum anomalous Hall effect}, because it can be viewed as being intrinsic to the material and the magnetic order.

Experimentally, magnetic order in topological insulators has been realized by doping with magnetic dopants such as Cr \cite{Liu:2012,Zhang:2012b,Kou:2013,Chang:2013,Kim:2016}, Fe~\cite{Chen:2010,Okada:2011,Wray:2011,Honolka:2012,Kim:2015}, Mn~\cite{Chen:2010,Hor:2010,Vobornik:2011,Henk:2012}, and V~\cite{Chang:2015}. Under favorable conditions, a mass term is induced on the surface, which opens a gap in the surface dispersion~\cite{Henk:2012} detectable by ARPES experiments~\cite{Chen:2010}. Both the d.c.\ Hall resistance~\cite{Chang:2013,Kou:2013,Chang:2015} and the Faraday- and Kerr rotation~\cite{Okada:2016} have been measured. The main drawback of magnetic doping is that the distribution of dopants and their magnetization is hard to control~\cite{Kim:2017}.

To obtain a surface magnetization that is stable at higher temperatures, proximity to a magnetically ordered layer has been proposed~\cite{Vobornik:2011,Eremeev:2013} and realized. Particularly promising layer materials are the insulating ferromagnet EuS~\cite{Wei:2013,Li:2015b} and the ferrimagnet yttrium iron garnet (YIG)~\cite{Lang:2014,Liu:2015a,Che:2018}. Even though EuS by itself favors an in-plane magnetization, coupling to the surface states of a topological insulator will lead to an out-of-plane magnetization~\cite{Semenov:2012}, realizing the mass term for Dirac fermions. Recent experiments with neutron scattering report that the magnetization at the interface even persists to room temperature~\cite{Katmis:2016,Kim:2017}, well above the Curie temperature of isolated EuS. In comparison, the advantage of yttrium iron garnet is that its Curie temperature is naturally above room temperature.

While introducing magnetic order on the surface would be sufficient to obtain the quantum anomalous Hall effect, a related approach is to consider magnetic order in the entire bulk, which some of the cited works already realize experimentally via bulk doping. Moreover, inspired by the proximity effect, one can consider topological insulators with and without magnetic order in heterostructures~\cite{Otrokov:2017,Xiao:2018}. Also, the magnetic ordering in the bulk need not be ferromagnetic: antiferromagnetism will also lead to a mass gap on the surface if there are parallel layers where all magnetic moments all point in the same direction~\cite{Zhang:2019d,Otrokov:2018,Gong:2019}. Conceptually, however, the introduction of bulk magnetization departs from topological insulators with time-reversal symmetry, and is better understood in terms of crystalline topological insulators with magnetic space groups~\cite{Mong:2010,Kruthoff:2017a,Watanabe:2018b}.

%%%%%%%%%%%%%%%%%%%%%%%%%%%%%%%%%%%%%%%%%%%%%%%%%%%%%%%%%%%%
\section{Time-reversal symmetric magnetoelectric effect}
We now return to the time-reversal symmetric topological insulator and describe a time-reversal-symmetric topological charge response. As explained before, it is necessarily nonlinear. We will consider a response not just to a magnetic field, but to a combined  electric and magnetic field.

We focus on the charge response, i.e.~the charge $Q(\E,\B)$ induced
by an external electric field $\E$ and magnetic field $\B$. Under
time reversal, the fields transform as $\E\to\E$, $\B\to-\B$, whereas
the charge stays invariant. Thus, the response is time-reversal-symmetric
if it satisfies $Q(\E,\B)=Q(\E,-\B)$. In contrast, the quantum Hall effect on the surface is not time-reversal-symmetric if the Hall conductivity is kept invariant under time-reversal, because the charge induced by a magnetic test field changes sign.

\begin{figure}[t]%
\begin{center}
  \includegraphics*[width=\linewidth]{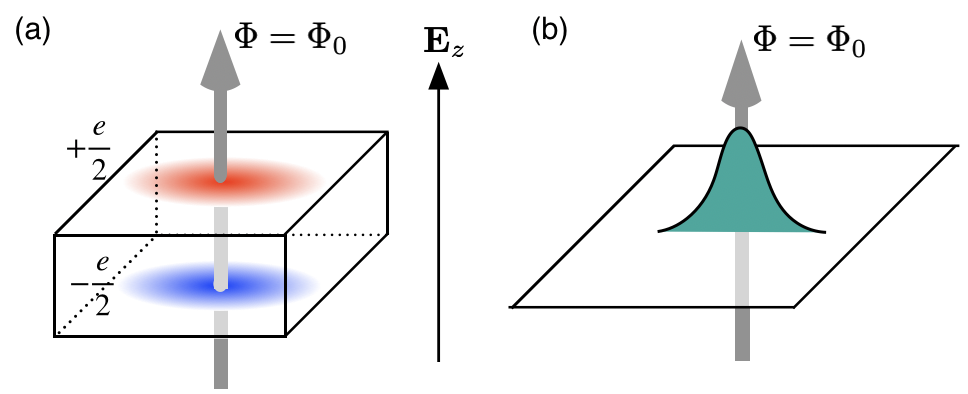}    
\end{center}
\caption{%
Time-reversal-symmetric nonlinear magnetoelectric effect.
(a) Topological insulator subject to a small electric field $\mathbf{E}_z$ in vertical direction, and threaded by a thin magnetic flux tube. When changing the flux $\Phi$ from zero to one flux quantum $\Phi_0=h/e$, a charge $Q = +(e/2)\mathop{\text{sgn}}(\mathbf{E}_z)$ will be transferred from the bottom to the top surface.
(b) Zero energy state bound to the flux tube on a single surface. It carries fractional charge $\pm e/2$. The electric fields controls whether it is occupied or unoccupied.
\label{fig:effect}}
\end{figure}

We consider a TI subject to a small uniform electric field, and threaded by a thin magnetic flux tube, both in vertical direction. To present the main idea, we first imagine a cubical geometry [see Fig.~\ref{fig:effect}].
We claim that when the flux is increased from zero, $\Phi=0$, to one flux quantum, $\Phi=\Phi_{0}\equiv h/e$, a half-integer charge will be transferred from the bottom to the top surface.
This effect can be understood as follows: The topological surface states are described by Dirac fermions. For Dirac fermions in an infinite two-dimensional plane, it has been shown that a magnetic flux will bind one (power law) localized  zero energy state per flux quantum~\cite{Aharonov:1979}.
Surprisingly, such a bound state carries a \emph{fractional} charge $\pm e/2$ compared to the reference configuration without flux\footnote{The chemical potential is chosen at zero energy, so that all states below zero energy are occupied.}~\cite{Jackiw:1981a,Krive:1987,Leinaas:2009,Leinaas:2009a}.
The charge has negative sign when the state is occupied by an electron, and positive sign when it is unoccupied. In a finite geometry, we expect two states, one on the bottom, and one on the top surface, where the flux tube pierces the material. At exactly zero energy, the occupation of the states would be ambiguous, but the small electric field lifts the degeneracy, and a charge $Q = +(e/2)\mathop{\text{sgn}}(\mathbf{E}_z)$ is induced on the top surface. This topological magnetoelectric effect is time-reversal-symmetric, but nonlinear.

We note that this effect is not related to the wormhole effect~\cite{Rosenberg:2010}, where it was reported that a magnetic flux tube inserted through a single lattice plaquette may open a conducting channel inside the bulk, but only if the diameter of the tube is much smaller than the lattice spacing of a tight-binding model. Here, we focus on a different situation where the flux tube is thin, but still covers many plaquettes, so that the low energy states of the TI are confined to the regular surfaces and can be described a two-dimensional continuum Dirac Hamiltonian.

\begin{figure}[t]%
\includegraphics*[width=\linewidth]{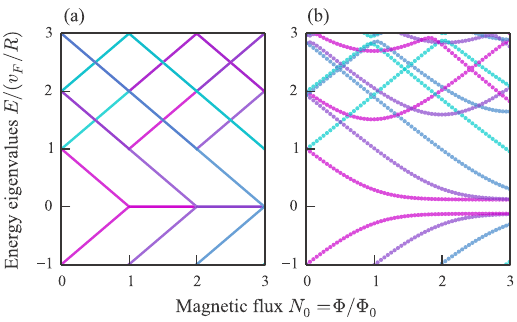}
\caption{%
Energy spectrum of the surface states of a TI with a magnetic flux tube~\cite{Zirnstein:2017}.
(a) Analytical results for a spherical geometry in the limit of a
thin flux tube. The colors encode angular momentum. For each flux quantum, one pair of states joins at zero energy.
(b) Numerical results for the low energy states of a finite lattice with cubic geometry. Formerly zero energy states are now separated by a small energy gap, which is due to the finite (relative) diameter of the flux tube.
\label{fig:spectrum}}
\end{figure}

To firmly establish that the flux tube also gives rise to zero energy states in a finite geometry, we have calculated the energies of the surface states analytically for a spherical surface~\cite{Zirnstein:2017}.
The surface states are modeled by the Dirac Hamiltonian on a sphere with radius $R$ \cite{Lee:2009,Imura:2012,Parente:2011}. In the presence of a magnetic flux tube and without electric field, it reads
\begin{align}
    \label{eq:hamiltonian}
    \hat H &= \frac{v_F}R \begin{pmatrix}
        0 & h^+ \\
        h^- & 0
    \end{pmatrix}
,\\ \nonumber
h^{\pm} &= \mp \left(\partial_\theta + \frac12 \cot \theta\right) + \frac{i\partial_\phi}{\sin \theta} + eR\mathbf{A}_\phi
,\end{align}
where $v_F$ denotes the Fermi velocity, $\theta$ and $\phi$ are spherical coordinates and $\mathbf{A}_\phi(\theta)$ is a magnetic vector potential of the flux tube with flux $\Phi$.
The eigenvalue problem $\hat H\psi=E\psi$ for the wave function $\psi(\theta,\phi)$ is simplified by introducing half-integer angular momentum $m=\pm \frac12,\pm \frac32,\dots$ and expressing the wave function as a product $\psi(\theta,\phi)=\tilde{\psi}(\theta)e^{im\phi}/\sqrt{R}$. Importantly, the boundary conditions are that $\tilde{\psi}(\theta)$ stays finite at the poles.
In the absence of an electric field, and in the limit of a thin flux
tube, the eigenenergies are~\cite{Zirnstein:2017}
\begin{equation}
    \label{eq-surface-energy-levels}
    E = \pm \begin{cases}
        c+n &\text{ if } c>0\\
        0 \text{ or } 1-c+n &\text{ if } c<0
    ,\end{cases}
,\end{equation}
where $c=(m+1/2-\Phi/\Phi_0)$ combines angular momentum $m$ and the magnetic flux. This spectrum is shown in Fig.~\ref{fig:spectrum}(a). At zero flux, the energies are separated by a finite size gap of order $v_{F}/R$. However, for each flux quantum, one pair of eigenstates joins at zero energy, and we can apply the previous argument, showing that one half-integer charge is transferred.

\begin{figure}[t]%
\includegraphics*[width=\linewidth]{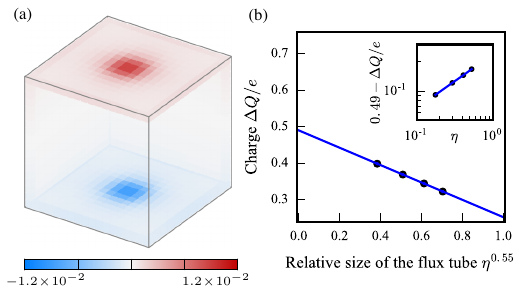}
\caption{%
Numerical calculation of the charge in a TI subject to a small electric field and a magnetic flux tube~\cite{Zirnstein:2017}.
(a) Charge distribution for a flux tube with finite diameter.
(b) Finite size scaling where the relative diameter $\eta$ of the flux tube shrinks to zero.
\label{fig:bulk}}
\end{figure}

Since this analytical calculation only applies in the idealized limit of a thin flux tube, we have also performed a numerical calculation on a lattice for the more realistic situation where the flux tube has a finite diameter. As shown in Fig.~\ref{fig:spectrum}(b), we now find that the states at one flux quantum have a small energy splitting, though it has to vanish as the flux tube shrinks.
Now, strictly speaking, only zero energy states carry half-integer charge, and it is not a priori clear how states with an energy splitting can yield fractional charge.
But we claim that the small electric field also helps with that: By choosing its energy $eR\mathbf{E}_z$ to be larger than the splitting, we can still polarize the occupation of the two states, but we also choose it smaller than the finite size gap $v_F/R$, so that the reference configuration without flux remains undisturbed. This choice is possible in a setup of mesoscopic size, where this finite size gap is appreciable.
We have verified numerically that such an electric field will indeed induce  a charge distribution with half-integer charge after threading a flux tube, by performing a finite size scaling where the size of a finite flux tube is shrunk to zero [see Fig.~\ref{fig:bulk}]~\cite{Zirnstein:2017}. This also shows that the bulk states do not contribute, a fact that cannot be assumed a priori, as we have seen in our discussion on anomaly matching.

\begin{figure}[t]%
\begin{center}
  \includegraphics*[width=1.0\linewidth]{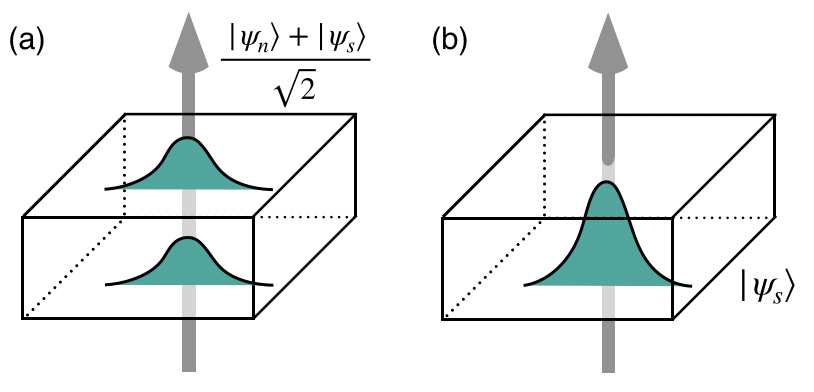}  
\end{center}
\caption{%
Illustration of surface states bound to a flux tube of one flux quantum in the presence of a small energy splitting.
(a) In the absence of an electric field, the symmetric superposition is occupied.
(b) In the presence of an electric field, the state, say, on the bottom surface is occupied. Relative to the previous state, its expected charge density differs by $\pm e/2$ on top and bottom surfaces.
\label{fig:states}}
\end{figure}

We now present an argument for how the states bound to the flux tube give rise to a half-integer charge, even when the states have a small energy splitting. We focus on lattice models, where the total charge is integer.
First, we note that when discussing charge, we refer to the expectation value of the charge density of a non-interacting many-electron system,
\begin{equation}
    \langle \rho(\mathbf{x}) \rangle
  =
    (-e)\sum_{E_\alpha < 0} |\eta_\alpha(\mathbf{x})|^2
,\end{equation}
where $E_\alpha$ denote the energies of the single-particle eigenfunctions $\eta_\alpha(\mathbf{x})$. 
While there is always an integer number of electrons in the system, and thus the total charge is an integer multiple of $(-e)$, the electron wave functions may be extended in space, leading to a \emph{local} charge imbalance with fractional values.
Second, in order to measure such an imbalance, we have to specify a reference configuration, i.e.~charge neutrality.
We choose the system at half-filling without electric field and with zero magnetic flux. In Appendix \ref{sec:fractional-charge}, we show that the expected charge density of this configuration is spatially uniform, provided there is an additional chiral pseudosymmetry, which can be realized e.g.~in bipartite lattices, and which we assume to be present here. Moreover, the density stays uniform for any value of the magnetic flux, because this does not affect the chiral pseudosymmetry.
However, at one flux quantum, a pair of eigenstates appears whose energy is almost zero, safe for a small splitting due to hybridization of the two bound states.
These eigenstates are the symmetric and antisymmetric linear combinations $(|\psi_{\text{n}}\rangle \pm |\psi_{\text{s}}\rangle)/\sqrt{2}$ of two states $|\psi_{\text{n}}\rangle$ and $|\psi_{\text{s}}\rangle$ that are localized on the top and bottom surfaces [see Fig.~\ref{fig:states}]. Due to the chiral pseudosymmetry, the energies come in pairs of positive and negative energies, and only the symmetric combination is occupied and contributes to the expected charge density, while the antisymmetric is unoccupied and does not contribute.
But if we consider a situation with a small electric field that is larger than the energy splitting, then the eigenstates will be $|\psi_{\text{n}}\rangle$ and $|\psi_{\text{s}}\rangle$, and, say, only the second one will be occupied and contribute to the charge density. Compared to the previous situation without electric field, this wave function contributes a local charge $-e/2$ on the top surface, and $+e/2$ on the bottom surface.
Similarly, increasing the flux from zero to one flux quantum in the presence of an electric field will also  give rise to a half-integer charge locally.

Observing this time-reversal symmetric charge response experimentally would require a mesoscopic setup and the realization of a thin magnetic flux tube. For the latter, type-II superconductors may be suitable, because they will allow a magnetic field only in quantized vortices. By pinning two superconducting vortices, or joining them in a giant vortex, as reported in Refs.~\cite{Kanda:2004,Cren:2011}, a tube with flux $\Phi_0=h/e$ could be established. Its diameter will be finite (reported $\sim 30\text{nm}$), but this is small compared to a mesoscopic system diameter of, say, 1$\mu$m, and hence would still lead to a sizable charge response, as the numerical results in Fig.~3(b) indicate. The limit of a very thin tube could be obtained by finite-size scaling of the measured data. The key difference between the response of a TI surface and an ordinary two-dimensional electron gas would be that the former leads to a half-integer multiple of the elementary charge, while the latter features an integer multiple.

%%%%%%%%%%%%%%%%%%%%%%%%%%%%%%%%%%%%%%%%%%%%%%%%%%%%%%%%%%%%
\section{Discussion}

We have scrutinized the magnetoelectric response of three dimensional time-reversal-symmetric TIs. The low-energy physics of such TIs is 
described by the axion action, and naively one might deduce the existence of a quantized linear magnetoelectric effect from it. We have discussed that for  periodic boundary conditions, the axion action does not contribute to the classical electromagnetic response at all, while for a finite system, it  contributes a boundary response. However, one also has to include the electromagnetic response  of the gapless boundary states, which cancels the boundary response from the axion action in order to preserve time-reversal symmetry. 
This cancellation is a manifestation of \emph{anomaly matching}, i.e.~the fact that quantum anomalous contributions from bulk and boundary have to cancel each other.  
Then, we have discussed the Witten effect, which tells us that an electric charge $\pm e\theta / 2\pi$ will be bound to a magnetic monopole due to the axion action; this can be considered a genuine bulk effect. But since magnetic monopoles are not available as elementary particles, this has no experimental consequences.
Instead, recent experiments have measured the quantum Hall effect of the surface electron gas, for which one has to explicitly break time-reversal symmetry, either by applying a strong magnetic field, by introducing magnetic order via magnetic dopants or proximity to an insulating ferromagnet.

In search for a time-reversal-symmetric topological effect, we have discussed a nonlinear magnetoelectric response. Here, the key result is that a thin magnetic flux tube will bind zero energy states on the surface, which carry half-integer charges relative to a reference state at half filling.  By utilizing a small electric field, which preserves time-reversal symmetry, the occupation of zero energy states can be controlled, and we obtain a local, experimentally observable, and quantized charge response of $Q = +(e/2)\mathop{\text{sgn}}(\mathbf{E}_z)$.

\begin{acknowledgement}
This research was funded by the German Research Foundation within the Collaborative Research Centre 762 (project B6).
\end{acknowledgement}

% Use the following code if you wish to generate your bibliography with BibTeX;
% replace the string "pss_demo" below with the name(s) of
% the BibTeX data base(s) you want to use.
% The resulting bibliography-output (the content of the .bbl file)
% must be pasted back into this file before submission.
% Please also include your BibTeX data base file(s) in your submission
% so that we can re-run BibTeX if necessary.
%
\bibliographystyle{pss}
\bibliography{magnetoelectric.bib}
%
% Replace the following example bibliography with your references
% before submission:

%%%%%%%%%%%%%%%%%%%%%%%%%%%%%%%%%%%%%%%%%%%%%%%%%%%%%%%%%%%%
\appendix
%%%%%%%%%%%%%%%%%%%%%%%%%%%%%%%%%%%%%%%%%%%%%%%%%%%%%%%%%%%%
\section{Action for classical electrodynamics}
\label{sec:electrodynamics}
In this appendix, we briefly review the action formalism for classical electrodynamics. We set $c=1$.

Since the electromagnetic fields $\E$ and $\B$ are not independent, one cannot immediately construct an action that reproduces all four Maxwell equations at once. Instead, one typically considers the Gauss law for the magnetic field and the law of induction as constraints, which are solved by introducing the electric potential $\phi$ and the magnetic vector potential $\mathbf{A}$, so that $\B = \nabla\times\mathbf{A}$ and $\E = -\nabla \phi - \partial_t \mathbf{A}$. Then, the Gauss law for the electric field and the Maxwell-Amp\`{e}re law can be obtained from an action principle by varying these fields.
Moreover, it is convenient to use relativistic notation, where the components of the gauge field $A_\mu$ are defined by $A_0=\phi$ and $A_i = -\mathbf{A}_i$ for $i=1,2,3$. The electric field components are neatly combined into the field tensor $F_{\mu\nu}=\partial_{\mu}A_{\nu}-\partial_{\nu}A_{\mu}$. Similarly, the four-current $j^\mu$ comprises the charge density $\rho$ and the current $\mathbf{j}$ via $j^0=\rho$, $j^i = \mathbf{j}_i$.
In this notation, the action for classical electrodynamics is given by
\begin{subequations}
\begin{align}
  S
  &=
  S_{\text{Maxwell}} + S_{\text{mat}}
\\
  S_{\text{Maxwell}}
  &=
  \int d^{3}xdt\, \left[-\frac{1}{16\pi}F_{\mu\nu}F^{\text{\ensuremath{\mu\nu}}}\right]
\\
  S_{\text{mat}}
  &=
  \int d^{3}xdt\, \left[- A_{\mu}j^{\mu}\right]
.\end{align}
\end{subequations}
The first term describes the dynamics of the electromagnetic field, while the second term couples matter (charge and matter) to the field. Evidently, the four-current can be obtained from the functional derivative $j^\mu = -\delta S_{\text{mat}} / \delta A_\mu$. In a material, these currents are themselves induced by the electromagnetic field. This material response can be captured by expressing the second term as a function of the field alone, $S_{\text{mat}}[A_\mu]$; this captures the kinetic and potential energy of the generated currents in the presence of the field. The currents can still be retrieved by the functional derivative. The axion action \eqref{eq:axion} is to be understood in this way, although for a realistic materials, which also show a dielectric response, one has to add other terms to obtain the full $S_{\text{mat}}$.

%%%%%%%%%%%%%%%%%%%%%%%%%%%%%%%%%%%%%%%%%%%%%%%%%%%%%%%%%%%%
\section{Fractional charge}
\label{sec:fractional-charge}
In this appendix, we show that in the presence of local chiral pseudosymmetry, the charge density of a lattice system of non-interacting electrons is spatially uniform at half filling if there are no states at exactly zero energy.~\cite{Zirnstein:2017}

We consider a finite lattice system where sites (unit cells) are labeled by $\mathbf{x}$ and have $n$ degrees of freedom. The dynamics of the electrons is described by a single-particle tight-binding Hamiltonian $\hat H$.
The chiral pseudosymmetry is given by a local operator $\hat C$ which anticommutes with the Hamiltonian, $\{\hat H,\hat C\}=0$.
For instance, for bipartite systems where hopping only occurs between sublattices, this operator can often be realized by multiplying the wave function on one sublattice with $(-1)$. Anticommutation implies that eigenstates come in pairs with positive and negative energies $E=E_\alpha=-E_\beta$, and locality means that their probabilities are equal $|\eta_\beta(\mathbf{x})|^2 = |\hat C\eta_\alpha(\mathbf{x})| = |\eta_\alpha(\mathbf{x})|^2$. In particular, we can express the probability as half the sum of a positive and a negative energy state, $|\eta_\alpha(\mathbf{x})|^2 = \frac12(|\eta_\alpha(\mathbf{x})|^2+|\eta_\beta(\mathbf{x})|^2)$. But this means that the expected charge density at half filling can be rewritten as a sum over all states, 
\begin{align}
  \langle \rho(\mathbf{x}) \rangle
&=
  (-e)\sum_{E_\alpha \leq 0} |\eta_\alpha(\mathbf{x})|^2 
\\ &=
  (-e)\sum_{E_\alpha \leq 0} \frac12(|\eta_\alpha(\mathbf{x})|^2 + |\eta_\beta(\mathbf{x})|^2)
\nonumber \\ &
  = (-e)\frac12 \sum_{E_\alpha} |\eta_\alpha(\mathbf{x})|^2
,\end{align}
if there are no states at exactly zero energy. However, since the eigenstates form a complete basis of the Hilbert space, this is constant
\begin{equation}
  \langle \rho(\mathbf{x}) \rangle = (-e)\frac{n}2
,\end{equation}
where $n$ counts the degrees of freedom per site (unit cell).

While the Hamiltonian $\hat H=\hat H(A_0,\mathbf{A})$ of the topological insulator also includes a coupling to the electric potential $A_0$ and the magnetic vector potential $\mathbf{A}_0$, this argument still applies as long as the pseudosymmetry $\hat C$ anticommutes with the Hamiltonian. For common lattice models, this is indeed the case even in the presence of the flux tube as long as the electric potential is zero, $A_0=0$. However, this implies that the charge density is unchanged, and that there is no charge response in this situation.

\end{document}